\documentclass[showpacs,aps,twocolumn]{revtex4}
\usepackage{bm}
\usepackage{amsmath}
\usepackage{graphicx}
\usepackage{subfigure}
\usepackage[usenames,dvipsnames]{color}
\definecolor{darkblue}{RGB}{0,0,196}
\usepackage[colorlinks=true,linkcolor=darkblue,citecolor=darkblue,urlcolor=darkblue]{hyperref}

\usepackage{setspace}
\usepackage{footmisc}
\usepackage[makeroom]{cancel}
\usepackage{comment}

\def\be{\begin{equation}}
\def\ee{\end{equation}}
\def\ba{\begin{eqnarray}}
\def\ea{\end{eqnarray}}

\usepackage{graphicx}
\usepackage{amsmath,bbm}
\usepackage{amssymb,bm}

\begin{document}

\title{Limiting Fragmentation in a Thermal Model with Flow}
\author{Swatantra~Kumar~Tiwari}
%\author{a}
%\author{Raghunath Sahoo}
%\author{b}
\author{Raghunath~Sahoo\footnote{Corresponding author: $Raghunath.Sahoo@cern.ch$}}
%\author{c}

\affiliation{Discipline of Physics, School of Basic Sciences, Indian Institute of Technology Indore, Indore- 453552, INDIA}

\begin{abstract}
\noindent
 The property of limiting fragmentation of various observables such as rapidity distributions ($dN/dy$), elliptic flow ($v_{2}$), average transverse momentum ($\langle p_{T} \rangle$) etc. of charged particles is observed when they are plotted as a function of rapidity ($y$) shifted by the beam rapidity ($y_{beam}$) for a wide range of energies from AGS to RHIC. Limiting fragmentation (LF) is a well studied phenomenon as observed in various collision energies and colliding systems experimentally. It is very interesting to verify this phenomenon theoretically. We study such a phenomenon for pion rapidity spectra using our hydrodynamic-like model where the collective flow is incorporated in a thermal model in the longitudinal direction. Our findings advocate the observation of extended longitudinal scaling in the rapidity spectra of pions from AGS to lower RHIC energies, while it is observed to be violated at top RHIC and LHC energies. Prediction of LF hypothesis for Pb+Pb collisions at $\sqrt{s_{NN}}$=5.02 TeV is given.
\end{abstract}

\maketitle
%

%%%%%%%%%%%%%%%%%%%%%%%%%%%%%%%%%%%%%%%%%%%%%%%%%%%%%%%%%%%%%%%%%%%%%%%%%%%%%%
\section{Introduction}
\noindent
%%%%%%%%%%%%%%%%%%%%%%%%%%%%%%%%%%%%%%%%%%%%%%%%%%%%%%%%%%%%%%%%%%%%%%%%%%%%%%
The ultimate goal of heavy-ion collisions is to study a phase transition from a hot, dense hadron gas (HG) to a deconfined and/or chiral symmetric phase of quarks and gluons called a quark-gluon plasma (QGP) \cite{Singh:1993,Singh:1992,Satz:2000,Muller:1995,Shuryak:1980}. By colliding heavy nuclei, a fireball with a large energy density extending over a sufficiently large space-time volume can be created so that an equilibrated quark-gluon plasma may be formed. However, experimental and theoretical investigations made so far reveal that it is indeed difficult to get an unambiguous evidence for QGP formation. It is very important to understand the dynamics of the collisions in order to suggest a unique signal for QGP. Such information can be obtained by analysing the properties of various particles emitted from various stages of the collisions. The dynamics of the hadronic system can be best studied via hadron yields, ratios, rapidity distributions and transverse mass spectra \cite{Letessier:2004}. In this article, we focus only on the rapidity distributions of particles. Various types of formulations have been used to study the rapidity spectra of hadrons \cite{Landau:1953,Bjorken:1983,Schnedermann:1993,Braun:1996,Braun:1995,Schnedermann:1992,Feng:2011,Feng:2009,Hirano:2002,Morita:2002,Manninen:2011,Bass:1998,Mayer:1997,Heinz:1999,Becattini:2007,Becattini1:2007,Biedron:2007,Broniowski:2008,Cleymans:2008,Broniowski:2001,Broniowski:2002}.

It is interesting to view the distribution of the multi-particle production in the rest frame of one of the colliding nuclei. Such distribution exhibits scaling independent of center-of-mass energy, $\sqrt{s_{NN}}$, in the fragmentation region. The slope of the $dN/d\eta(y)$ curve in the fragmentation region of the projectiles (high $\eta(y)$ region) remains independent of $\sqrt{s_{NN}}$ for a given collision centrality. It was proposed by Benecke $et\; al.$ \cite{Benecke:1969}, Chou $et\; al.$ \cite{Chou:1970}, Feynman \cite{Feynman:1969}, and Hagedorn \cite{Hagedorn:1970} that as $\sqrt{s_{NN}}\rightarrow \infty$, the multiplicity distribution becomes independent of $\sqrt{s_{NN}}$. Here, the particle multiplicity refers to any of the secondaries produced out of the collision. This universality of multi-particle production is called limiting fragmentation. In a microscopic picture, while describing the system formed in high-energy collisions, the application of perturbative Quantum Chromodynamics (pQCD) refers to very high mean free path of the system quanta. On the other hand, the application of the statistical hadron gas model (SHGM) and hydrodynamic models requires small mean free path, barring the freeze-out hypersurface, where the mean free path is assumed to be higher than the system size. In the latter scenario of a picture of hydrodynamic evolution of the system following Landau hydrodynamics, the hypothesis of limiting fragmentation appears as a coincidence, where the particle multiplicity distribution follows a Gaussian (pseudo)rapidity profile. The use of a dynamical SHGM thus inherits the microscopic ingredient of a hydrodynamic evolution of the system to look into the possible hypothesis of limiting fragmentation. In the macroscopic picture, the Lorentz contracted volume $Vm_p/\sqrt{s}$, in hadronic and nuclear collisions controls the $dN/d\eta(y)$, the entropy of the system at freeze-out. Hence the final state (pseudo)rapidity distribution is a manifestation of the Lorentz contraction factor and thus, the maximum (pseudo)rapidity ($\equiv y_{beam}=ln(\frac{\sqrt{s_{NN}}}{m_p})$, the beam rapidity, where $m_p$ is the mass of the proton) achieved at a given collision energy. Usually one studies the pion $dN/d\eta(y)$ for their maximum production probability in a multiparticle production process.

In collision experiments, limiting fragmentation is a much discussed phenomenon because it is observed for various colliding systems {\it e.\;g.} $e^{+} +  e^-$ \cite{Back:2006}, $p+p(\bar{p})$ ($\sqrt{s}$ = 53 GeV to 900 GeV for charged particles in $|\eta-y_{beam}| > -2.5$ )\cite{UA5}, d+Au, Au+Au etc. \cite{Back:2006,Alver:2009,Adams:2005aa,Adams:2005cy,Abelev:2010,BackR:2005,phobos-prl-91}. The limiting fragmentation or the longitudinal scaling, as it is named otherwise, has also been observed for photons produced in forward rapidities in heavy-ion collisions (Au+Au) at RHIC energies \cite{Adams:2005aa,Adams:2005cy}. In a recent work of LHCf Collaboration \cite{Adriani:2015iwv}, the inclusive production of $\pi^0$ is measured in p+p collisions at 2.76 TeV and 7 TeV and in p+Pb collisions at 7 TeV energies at a very forward region $(8.8 < y < 10.8)$. In this work, the hypothesis of limiting fragmentation is found to be valid in p+p collisions. However, for the $p+p$ collisions at LHC $\sqrt{s}$ = 0.9, 2.76 and 7 TeV, the limiting fragmentation is reportedly violated in the measurements done for photons in the forward rapidity region $(2.3 < \eta < 3.9)$ \cite{ALICE:2014rma}. Although, the observed photons are decayed photons, primarily from $\pi^0$, because of adversely different rapidity coverage of ALICE PMD and LHCf, no conclusions could be made on the observation of limiting fragmentation in $\pi^0$ versus the violations in photons. 

Recently, various types of theoretical approaches have been used to study the limiting fragmentation phenomenon \cite{bialas,Ruan:2010,Nasim:2011,Gelis:2006,Bleibel:2016,Stasto:2011zza} observed in heavy-ion collisions. In Ref. \cite{bialas}, a two step process is used for soft particle production in hadronic collisions, where the first step is related to the multiple gluon exchange between the partons from the two colliding hadrons, while in the second step, partons radiate hadronic clusters. This two step mechanism explains the observation of limiting fragmentation in the rapidity spectra of particles in heavy-ion collisions. In Ref. \cite{Ruan:2010}, the limiting fragmentation and its possible violation is discussed within the partonic approach. Phenomenologically, limiting fragmentation suggests that the hadronic cross sections become independent of collisions energies. This means that the excitation and breakup of hadrons would be independent of collision energies and distributions in the fragmentation region would approach a limiting curve \cite{Ruan:2010}. But we know that the hadronic cross sections are not constant at very high energies \cite{Gelis:2006,atlas-cross}. Therefore, the limiting fragmentation should fail at such high energies. However, limiting fragmentation has been observed in a wider region at RHIC energies and it is referred to as an extended longitudinal scaling \cite{Back:2005}. In view of partons, longitudinal scaling relates to Bjorken scaling of the parton distributions and the production dynamics \cite{Ruan:2010}. Limiting fragmentation of $dN_{ch}/d\eta$, $v_2$ and $\langle p_T \rangle$ is studied by using transport models like, Ultra-relativistic Quantum Molecular Dynamics (UrQMD) and A Multi Phase Transport  (AMPT) at various $\sqrt{s_{NN}}$ \cite{Nasim:2011}. In ref. \cite{Nasim:2011}, it is observed that AMPT with the string melting scenario shows the longitudinal scaling for $dN_{ch}/d\eta$, $v_2$ and $\langle p_T \rangle$ while UrQMD and AMPT default versions show it only for $dN_{ch}/d\eta$ and $\langle p_T \rangle$. The authors argue that the longitudinal scaling in $dN_{ch}/d\eta$ and $\langle p_T \rangle$ does not necessarily mean that the same scaling will be observed in $v_2$. Sarkisyan {\it et al.} \cite{Edward:2016} have also presented the phenomenon of limiting fragmentation of pseudorapidity distributions for the charged particles in the framework of effective energy approach combined with the constituent quark picture with the use of Landau hydrodynamics. Cleymans {\it et al.} \cite{Cleymans:2008} have recently studied the extended longitudinal scaling for $dN/dy$ of pions using statistical thermal model. Their model assumes a Gaussian distribution of fireballs centered at zero rapidity, with a Boltzmann-like thermal single particle distribution. They fit the experimental data on pion rapidity spectra from SPS to RHIC energies with such Gaussian distributions and extract the fit parameters. After that, they extrapolate the fit parameters at LHC energy and predict the rapidity spectra for pions at this energy. They claim that the property of extended longitudinal scaling of rapidity spectra of pions is consistent with the statistical thermal model up to highest RHIC energies and it is violated at LHC. In Ref. \cite{Bleibel:2016}, the property of limiting fragmentation for $dN/dy$ for $p+p$ collisions is studied via Monte Carlo quark-gluon string model. It is found that the extended longitudinal scaling is also valid at LHC within this approach. Stasto \cite{Stasto:2011zza} has studied the property of limiting fragmentation for pseudorapidity distributions in nucleon- nucleon collisions for charged particles using the framework of $k_t$ factorization with unintegrated gluon distributions. In this work, the limiting fragmentation is observed for all energies taken into consideration. Furthermore, the calculations based on hadronic interactions models like DPMJET \cite{Bopp:2005cr,Engel:1996yb} and QGSJET \cite{Ostapchenko:2004ss} claim that the limiting fragmentation phenomenon in the rapidity distributions of $\pi^0$ is observed in p+p collisions at $\sqrt{s}$= 2.76 TeV and 7 TeV.

We plan to study the well established property of limiting fragmentation for rapidity distributions of pions in heavy-ion collisions using the statistical thermal model with the effect of flow \cite{Tiwari:2013}. For this purpose, we use our recently proposed excluded-volume model with the incorporation of collective flow in the longitudinal direction. This model has been successful in explaining various aspects of particle production in heavy-ion collisions, like- particle spectra, ratios etc. at RHIC and LHC energies \cite{Tiwari:2012}. We take the hadrons and their resonances having masses up to 2 GeV. We assign an equal hard-core size to each type of baryons in the hadron gas (HG) in order to include repulsive interactions between them, while the mesons, which can interpenetrate into each other are treated as point-like particles. We impose the strangeness neutrality condition, $\sum_{i}S_{i}(n_{i}^{s}-\bar{n}_{i}^{s})=0$, where $S_{i}$ is the strangeness of the $i-th$ hadron in order to ensure the strangeness conservation in our model. We use the chemical freeze-out criteria proposed in our model \cite{Tiwari:2012} to obtain temperature (T) and baryon chemical potential ($\mu_B$) at various center-of-mass energies. The paper is organized as follows: we first discuss the formulation of our model for HG and then we discuss its applicability in describing the rapidity distribution. After that, we modify our thermal model by incorporating the collective flow in the longitudinal direction. In the ensuing section, we compare the experimental data on various hadron ratios at LHC energy (2.76\;TeV) with our model predictions. We also deduce $dV/dy$ for pions at various $\sqrt{s_{NN}}$. Then, we calculate rapidity distributions for pions at various $\sqrt{s_{NN}}$ and shift the rapidity distributions in the rest frame of one of the beams. Finally, we present summary and conclusions. 

\section{The Model}

We have recently proposed an excluded-volume model for a hot and dense hadron gas \cite{Tiwari:2012} where we derive the number density $n_i^{ex}$ for the $i-th$ species of baryons using quantum statistics in the grand canonical partition function which is given after excluded-volume correction as follows \cite{Tiwari:2012} :
\begin{equation}
n_i^{ex} = (1-R)I_i\lambda_i-I_i\lambda_i^2\frac{\partial{R}}{\partial{\lambda_i}}+\lambda_i^2(1-R)I_i^{'},
\end{equation}
where $\displaystyle R=\sum_in_i^{ex}V_i^0$ is the fractional occupied volume by the baryons \cite{Tiwari:2013wga}. $\displaystyle V_i^0= 4\pi\;r^3/3$ is the eigen-volume of each baryon having a hard-core radius $r$ and $\lambda_i$ is the fugacity of the $i-th$ baryon. Here we take $r$=0.8\;$fm$ as a free parameter in our calculation. Further, $I_i$ is the integral of the baryon distribution function over the momentum space \cite{Tiwari:2012}. Here, we use quantum statistics in the grand canonical partition function for baryons. Since in this work we calculate the rapidity distributions only at LHC energies hence for the sake of convenience we use Boltzmann's statistics in the grand partition function for HG. Now in the Boltzmann's limit, the eq (1) can be reduced in the following form \cite{Mishra:2008tc} :
\begin{equation}
n_i^{ex} = (1-R)J_i\lambda_i-J_i\lambda_i^2\frac{\partial{R}}{\partial{\lambda_i}},
\end{equation}
where $J_i$ is the momentum integral for baryons in the Boltzmann statistics. Now, eq. (2) can be rewritten as follows \cite{Tiwari:2013} :

\begin{widetext}
\ba
\frac{dN_i}{dy\;m_T\;dm_T\;d\phi}&=&\frac{g_iV\lambda_i}{(2\pi)^3}\;\Big[\Big((1-R)-\lambda_i\frac{\partial{R}}{\partial{\lambda_i}}\Big)\; \frac{E_i}{\displaystyle \Big[\exp\left(\frac{E_i}{T}\right)\Big]}\Big].
\ea
\end{widetext}
%\begin{widetext}
%\ba
%\frac{dN_i}{dy\;m_T\;dm_T\;d\phi}&=&\frac{g_iV\lambda_i}{(2\pi)^3}\;\Big[\Big((1-R)-\lambda_i\frac{\partial{R}}{\partial{\lambda_i}}\Big)\; \frac{E_i}{\displaystyle \Big[exp\left(\frac{E_i}{T}\right)+\lambda_i\Big]}
%\nonumber \\
%&&-\lambda_i(1-R)\;\frac{E_i}{\displaystyle \Big[exp\left(\frac{E_i}{T}\right)+\lambda_i\Big]^2}\Big],
%\ea
%\end{widetext}
Here $y$ is the rapidity variable and $m_T=\sqrt{{m}^2+{p_T}^2}$ is the transverse mass. $E_i$ is the energy of the $i-th$ baryon, $V$ is the total volume of the fireball formed at chemical freeze-out and $N_i$ is the total number of $i-th$ baryons. We assume that the freeze-out volume of the fireball for all types of hadrons at the time of the homogeneous emissions of hadrons remains the same. 

By using $E_i=m_T{\cosh}y$ in eq. (3) and integrating the whole expression over transverse component we get the rapidity distributions of baryons as follows \cite{Tiwari:2013} : 
\begin{widetext}
\ba
\Big(\frac{dN_i}{dy}\Big)_{th}&=&\frac{g_iV\lambda_i}{(2{\pi}^2)}\;\Big[\Big((1-R)-\lambda_i\frac{\partial{R}}{\partial{\lambda_i}}\Big)\; \int \frac{m_T^2\;{\cosh}y\;dm_T}{\displaystyle\Big[\exp\left(\frac{m_T\;{\cosh}y}{T}\right)\Big]}\Big].
\ea
\end{widetext}
Eq.(4) gives the rapidity distributions of baryons arising due to a stationary thermal source. Similarly, the rapidity density of mesons can be calculated by using the following formula \cite{Tiwari:2013} :
\begin{equation}
\Big(\frac{dN_m}{dy}\Big)_{th}=\frac{g_mV\lambda_m}{(2{\pi}^2)}\;\int \frac{m_T^2\;{\cosh}y\;dm_T}{\displaystyle\Big[\exp\left(\frac{m_T\;{\cosh}y}{T}\right)\Big]}.
\end{equation}
Here $g_m$, $\lambda_m$ are the degeneracy factor and fugacity of the meson $m$, respectively. Further simplifying the eq. (4) by integrating it from $m_T=m_i$ to $m_T=\infty$, we get the rapidity distribution of baryons in Boltzmann's statistics as follows \cite{Schnedermann:1993,Tiwari:2013pva} :

\begin{widetext}
\ba
\Big(\frac{dN_i}{dy}\Big)_{th}=\frac{g_iV\lambda_i}{2\pi^2}\;\Big[(1-R)-\lambda_i\frac{\partial{R}}{\partial{\lambda_i}}\Big]
\exp\left(\frac{-m_i\;{\cosh}y}{T}\right)\Big[m_i^2T+\frac{2m_iT^2}{{\cosh}y}+\frac{2T^3}{{\cosh}^2y}\Big].
\ea
\end{widetext}
In a similar fashion we can also find the formula for rapidity distribution for mesons in Boltzmann's statistics using our model as below :

\begin{widetext}
\ba
\Big(\frac{dN_m}{dy}\Big)_{th}=\frac{g_mV\lambda_m}{2\pi^2}\;\exp\left(\frac{-m_m\;{\cosh}y}{T}\right)\Big[m_m^2T+\frac{2m_mT^2}{{\cosh}y}+\frac{2T^3}{{\cosh}^2y}\Big],
\ea
\end{widetext}
where $m_m$ is the mass of the $m-th$ meson.

When we compare our model results with the experimental data on rapidity distributions of hadrons, we find that our model describes the experimental data very well at mid-rapidity but it fails at forward and backward rapidities \cite{Tiwari:2013}. Hence, we modify the expression for rapidity spectra for hadrons as obtained in our thermal model by incorporating a flow velocity in the longitudinal direction. Compared to the static fireball approximation used in the framework of statistical hadron gas models in describing heavy-ion collisions,  the inclusion of flow brings up the dynamical aspects, while describing the experimental data. While incorporating longitudinal flow in a stationary thermal source, a boost invariance scenario is modified by restricting the boost angle, $\eta$ to a fixed interval \cite{Schnedermann:1993}. Thus the resulting rapidity spectrum of the $i-th$ hadron, after the incorporation of the flow velocity in the longitudinal direction is \cite{Tiwari:2013} : 
\begin{eqnarray}
\frac{dN_i}{dy}=\int_{-\eta_{max.}}^{\eta_{max.}} \Big(\frac{dN_i}{dy}\Big)_{th}(y-\eta)\;d\eta,
\end{eqnarray}
where $\displaystyle\Big(\frac{dN_i}{dy}\Big)_{th}$ can be calculated by using eq. (6) for the baryons and eq.(7) for the mesons. The average longitudinal velocity is given as \cite{Tiwari:2013,Netrakanti:2005} :
\begin{eqnarray}
\langle\beta_L\rangle=tanh\Big(\frac{\eta_{max.}}{2}\Big).
\end{eqnarray}
Here $\eta_{max.}$ is a parameter which is used to provide the upper rapidity limit for the longitudinal flow velocity at particular $\sqrt{s_{NN}}$. The value of $\eta_{max.}$ is found to increase with $\sqrt{s_{NN}}$ and hence $\beta_L$ also increases.

\section{Results and Discussion}

\begin{figure}
\includegraphics[height=20em]{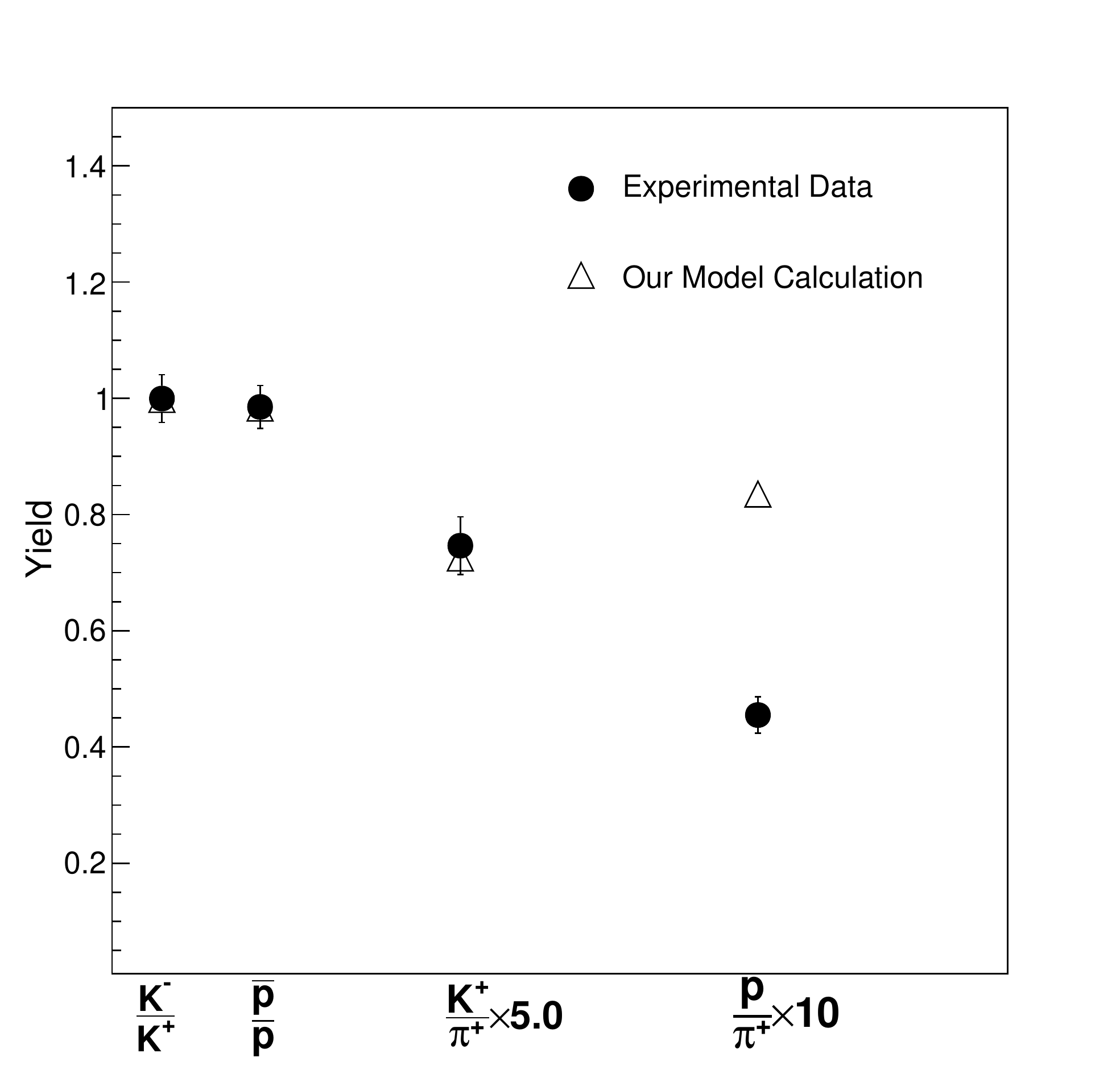}%use pdflatex for pdf
\caption[]{Various particle ratios at $\sqrt{s_{NN}}$=2.76\;TeV for the most central Pb-Pb collisions. Solid symbols are experimental data and open symbols are results of our model calculation.}
\end{figure}

In figure 1, we have shown the multiplicity of various particle ratios at center-of-mass energy of 2.76\;TeV. We have compared our model calculation with the experimental data \cite{Abelev:2012wca}. To calculate particle ratios at LHC energies we use the chemical freeze-out criteria as proposed in our model \cite{Tiwari:2012} from which we extract chemical freeze-out parameters {\it i.e.} temperature (T) and baryon chemical potential ($\mu_B$) at this energy. The values of T and $\mu_B$ at this energy are 163.5 MeV and 1.525 MeV, respectively which go in line with the observations by the ALICE experiment at LHC \cite{Abelev:2012wca}. We have also included the contributions of resonance decays while calculating the particle ratios. We find a very good agreement between our results and the experimental data except in the case of $p/\pi^+$ ratio, where our model result lies well above the experimental data. The thermal model fails to explain this ratio, which insights a new kind of formation mechanism. Various mechanisms have been used to explain this non-thermal particle ratio \cite{Steinheimer:2013,Noronha:2014}. In Ref. \cite{Steinheimer:2013}, it has been suggested that the $p/\pi^+$ ratio is strongly modified due to the late stage hadronic effects in which hadrons fall out of equilibrium until they finally freeze-out. Noronha-Hostler {\it et al.} \cite{Noronha:2014} pointed out that the inclusion of extended mass spectrum i. e. Hagedorn states, into the HG equation of state can explain such a suppressed $p/\pi^+$ ratio at LHC. The inclusion of baryon-antibaryon, $B-\bar{B}$ channels in hydrodynamical models reduce the final state proton and anti-proton multiplicity and hence explain the $p/\pi^+$ ratio at LHC, which is termed as {\it ``proton puzzle"} \cite{Song-Bass}.

\begin{figure}
\includegraphics[height=20em]{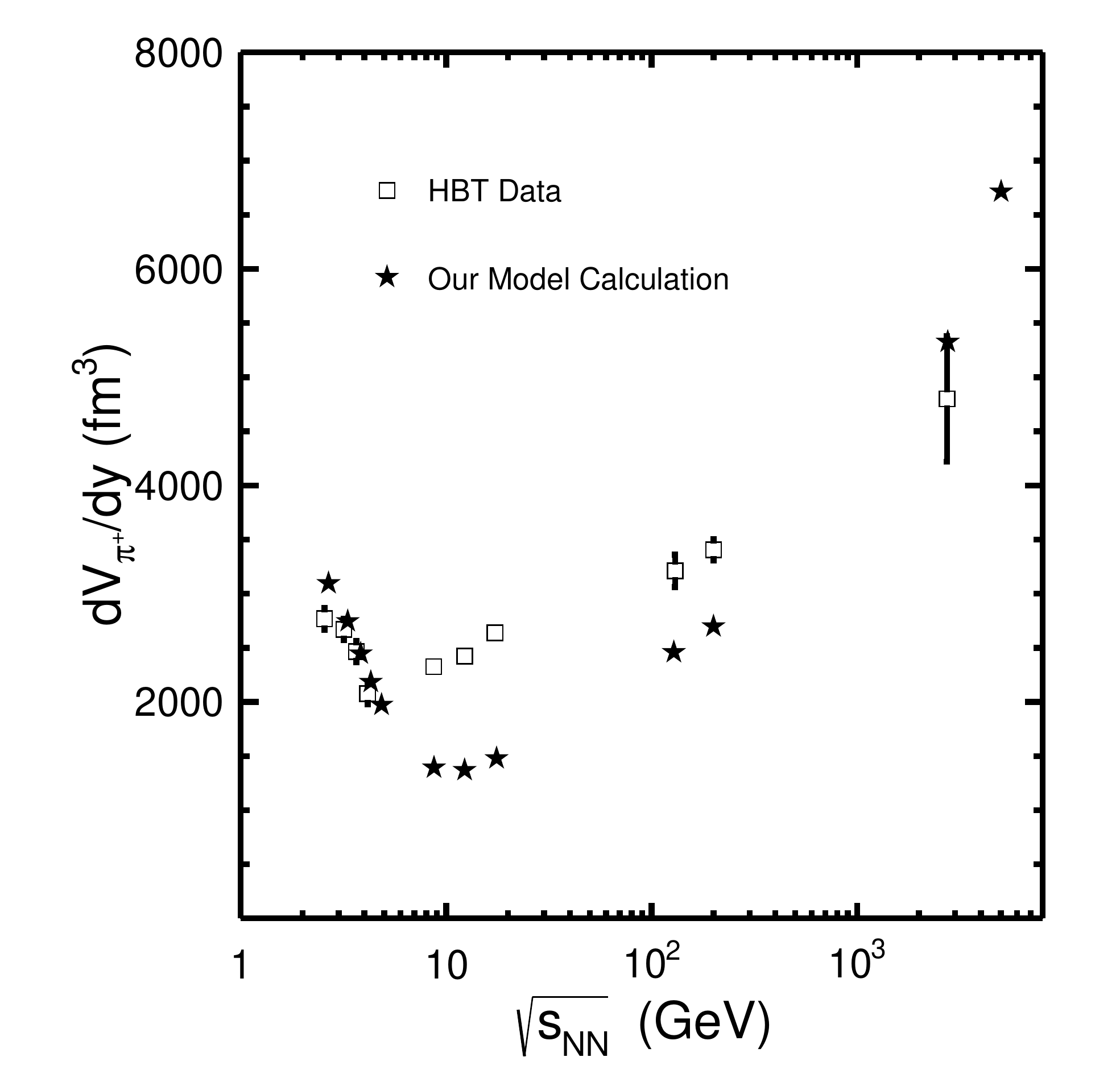}%use pdflatex for pdf
\caption[]{Energy dependence of freeze-out volume for central nucleus-nucleus collisions. Open symbols are HBT data points for $\pi^{+}$ and solid symbols are those calculated in our model.}
\end{figure}

\begin{figure}
\includegraphics[height=20em]{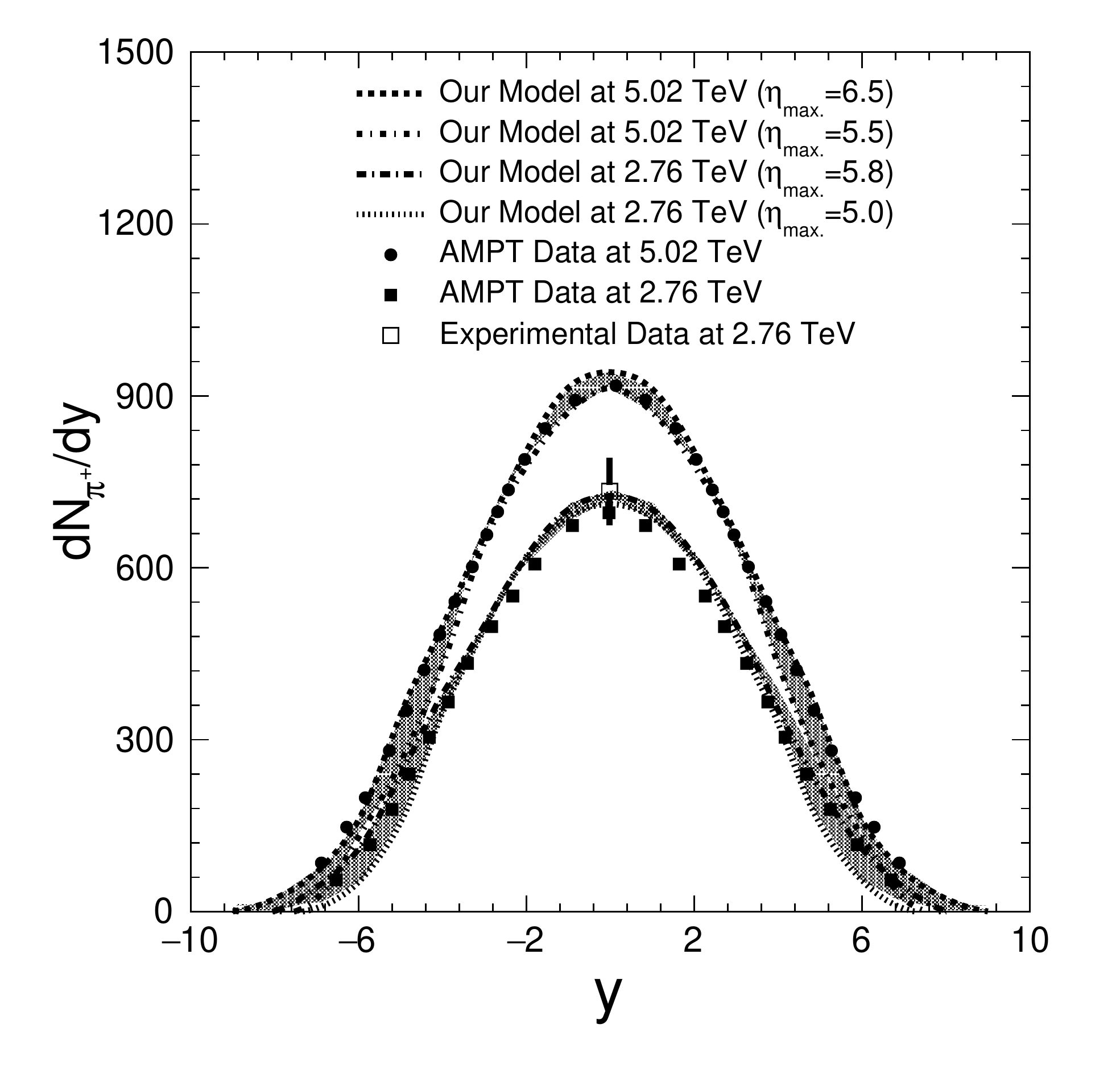}%use pdflatex for pdf
\caption[]{Rapidity distributions of $\pi^+$ at LHC energies. Lines are our model calculations and symbols are the results from AMPT model \cite{Ma:2016fve}. The only available experimental data at y=0 for $\sqrt{s_{NN}}$= 2.76 TeV Pb+Pb collisions is shown for a comparison. The shaded area shows the uncertainty arising due to variations of the parameter chosen for calculating rapidity distributions.}
\end{figure}

A statistical thermal model essentially describes the system in a thermodynamic equilibrium but it does not provide any information pertinent to the existence of a QGP phase before hadronization. However, if a mixed phase occurs in the space-time evolution of the system formed in heavy-ion collisions, the volume $V$ of the system at freeze-out is expected to be much larger than what we expect from a system if it remains only in the hadronic phase throughout the evolution. Figure 2 shows the variation of $dV/dy$ for $\pi^+$ as calculated in our model and their variations with the center-of-mass energy. To deduce $dV/dy$ for $\pi^+$, we use the experimental data for $dN/dy$ at mid-rapidity and divide it by the corresponding number density calculated in our model. At $\sqrt{s_{NN}}$=2.76\;TeV we have taken the experimental mid-rapidity data on $dN/dy$ for $\pi^+$ from Ref. \cite{Abelev:2013vea}. Since at $\sqrt{s_{NN}}$=5.02\;TeV there are no experimental data on the rapidity density of $\pi^+$ at mid-rapidity, we have used the AMPT data \cite{Ma:2016fve} at this energy to extract $dV/dy$ for $\pi^+$. This is because AMPT-SM data well describes the rapidity density at various collision energies \cite{Nayak:2016}. We extend our earlier studies \cite{Tiwari:2013} on mid-rapidity multiplicity distribution of pions at RHIC to the highest LHC energy through $\sqrt{s_{NN}}$=2.76\;TeV. We have compared our model predictions with the data obtained from the pion interferometry (HBT) \cite{Adamova:2003,Braun-Munzinger:2014lba}, which reveals the thermal (kinetic) freeze-out volume. 
After observing a reasonable agreement of the model calculations with that of experimental data on hadron yield ratios and studying $dV/dy$, we proceed to look into the hypothesis of limiting fragmentation and its possible validity at LHC energies.

\begin{figure}
\includegraphics[height=20em]{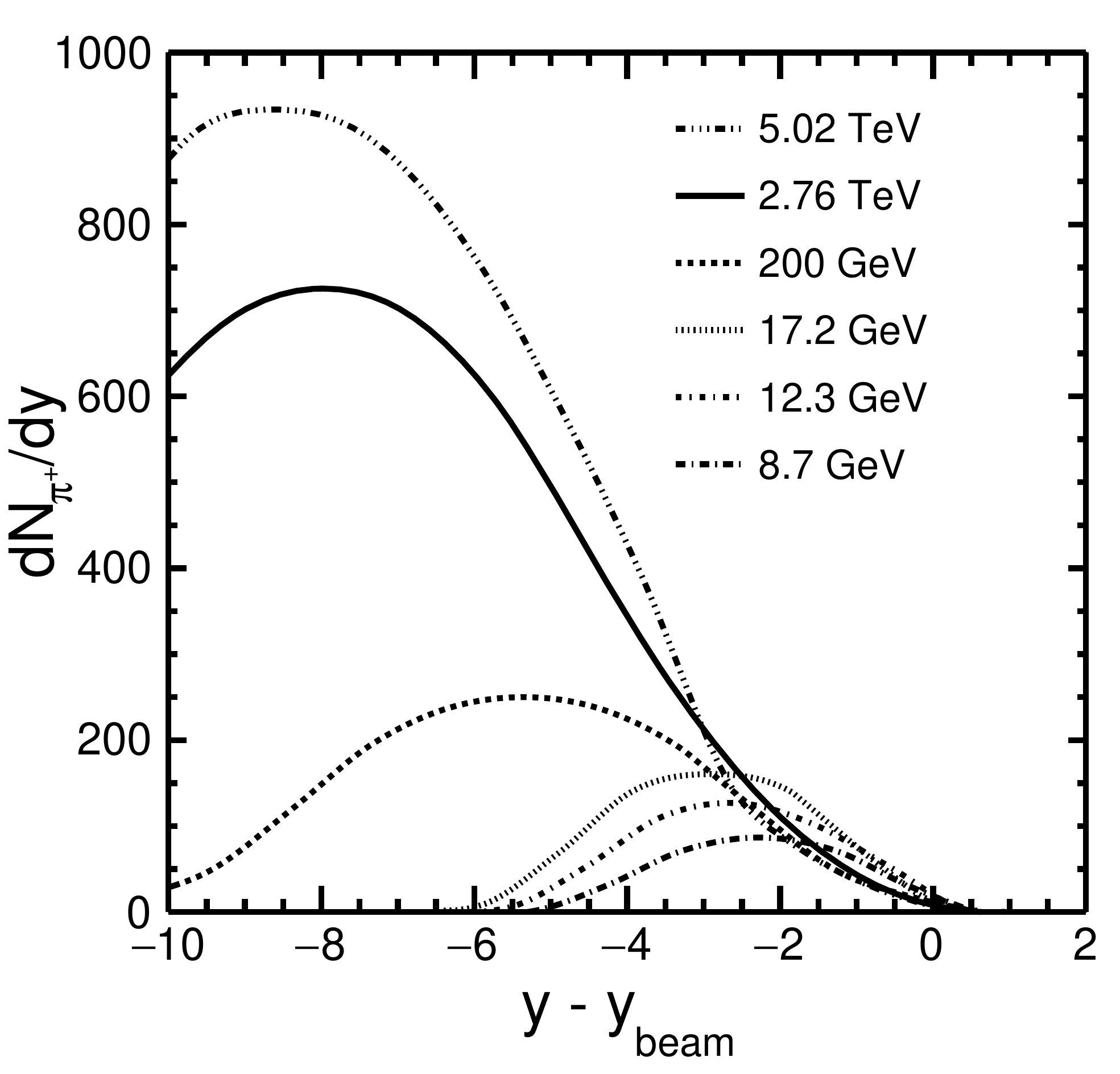}%use pdflatex for pdf
\caption[]{Limiting fragmentation as observed in rapidity distributions at various center-of-mass energies.}
\end{figure}

In figure 3, we show the rapidity distributions of $\pi^+$ for Pb+Pb collisions at 2.76 and 5.02 TeV energies at LHC. Due to lack of the experimental data on $\pi^+$ rapidity density over all the rapidities, we compare our model results with that of AMPT \cite{Ma:2016fve}, which uses the string melting scenario. Since, the AMPT model with string melting scenario is very successful in explaining the experimental data on charged particle rapidity distributions \cite{Nayak:2016}, it is reliable to compare our model results with that of the AMPT with string melting scenario, as far as the particle multiplicity density distribution is concerned. To calculate rapidity distributions, we use eq. (8), where we take the value of the parameter $\eta_{max.}$=5.8 at 2.76 TeV. In order to check the appropriateness of the  parameter, we also show the rapidity distributions with a slightly different value $\eta_{max.}$=5.0 at this energy. We find that $\eta_{max.}$ = 5.8 explains the data successfully. The shaded area shows the difference in rapidity distributions arising due to various values of $\eta_{max.}$. After using $\eta_{max.}$=5.8 in eq. (9), we get the longitudinal flow velocity, $\beta_L$=0.993$c$, where $c$ is the of speed of light. With this longitudinal flow incorporated in our thermal model, we get a reasonable agreement of $dN_{\pi^+}/dy$ between AMPT data and our model. We also show the experimental data available at mid-rapidity at $\sqrt{s_{NN}}$= 2.76 TeV \cite{Abelev:2013vea} and we observe that our model describes it very well. In the same way, we have taken $\eta_{max.}$=6.5 at $\sqrt{s_{NN}}$= 5.02 TeV. For comparison, we show the rapidity distributions for $\eta_{max.}$=5.5 at this energy. After comparison we see that $\eta_{max.}$=6.5 is appropriate in order to describe the rapidity distributions at this energy. Also, the shaded area describes the uncertainty in the rapidity distributions due to the use of different values of $\eta_{max.}$. This value of $\eta_{max.}$=6.5 gives the longitudinal flow velocity $\beta_L=0.996c$, which is almost the same as observed at $\sqrt{s_{NN}}$=2.76 TeV. Again, we observe a good agreement between our model calculations and the results as observed in the AMPT model \cite{Ma:2016fve} at $\sqrt{s_{NN}}$=5.02\;TeV. This comparison strengthens the appropriateness of the parameter $\eta_{max.}$ chosen in our model calculations and provides the more realistic value of longitudinal flow velocity ($\beta_L$).      

Figure 4 represents the variation of rapidity distributions of $\pi^+$ with respect to the shifted rapidity $y-y_{beam}$ over a broad energy range from 8.7 GeV to 5.02 TeV. Here, $y_{beam}$ can be calculated by using the formula, $y_{beam}=ln(\frac{\sqrt{s_{NN}}}{m_p})$ at each energy. In this work, we have calculated rapidity distributions at various energies using the discussed model with flow \cite{Tiwari:2013}. We find that the property of extended longitudinal scaling is observed up to below top RHIC energy, while it is observed to be violated at LHC energies in our thermal model with flow. In figure 5, we have shown the variation of the longitudinal flow extracted in our model with respect to the center-of-mass energy. We find that the longitudinal flow increases with the collision energy. 

\begin{figure}
\includegraphics[height=20em]{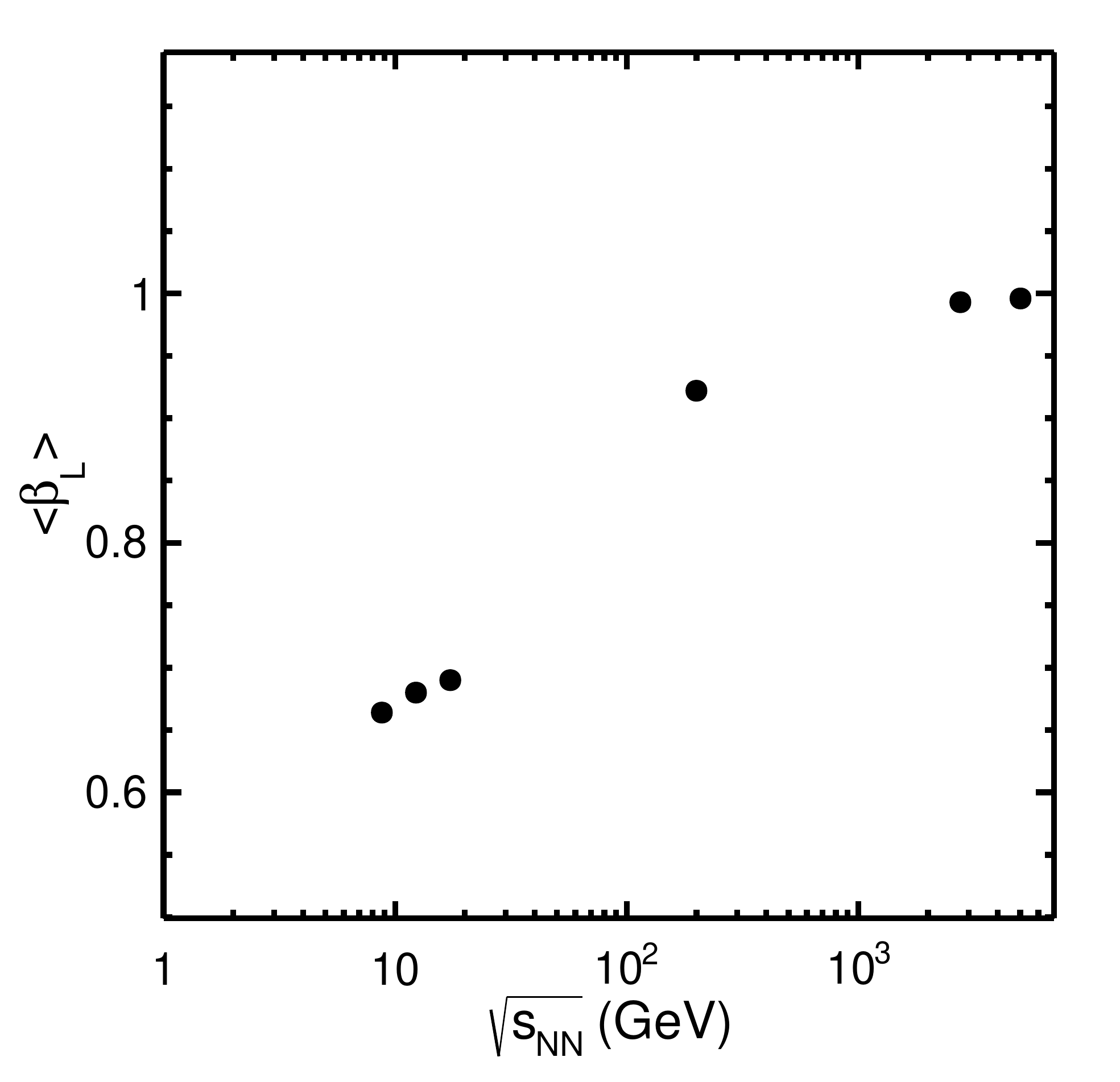}%use pdflatex for pdf
\caption[]{Variation of the longitudinal flow velocity (in units of $c$) with respect to $\sqrt{s_{NN}}$.}
\end{figure}

%%%%%%%%%%%%%%%%%%%%%%%%%%%%%%%%%%%%%%%%%%%%%%%%%%%%%%%%%%%%%%%%%%%%%%%%%%%%%
\section{Summary and Conclusion}
 We give a detailed overview of the experimental and theoretical findings related to the hypothesis of limiting fragmentation observed in various collision species, which is a very important phenomenon in multi-particle production processes. To see its validity at RHIC and LHC energies in the framework of a statistical hadron gas model, which has been very successful in describing various aspects of particle production and freeze-out in heavy-ion collisions, we incorporate flow to make the formalism more realistic. In this framework, we find that our model provides a good fit to the various particle ratios at LHC energy except $p/\pi^+$ ratio, which can be explained by some other kinds of production mechanism. We have also calculated the volume of the fireball at chemical freeze-out and we get a very large volume of the fireball at freeze-out, which suggests that there should be a mixed phase in the space-time evolution of the fireball formed in heavy-ion collisions. We have shown the rapidity spectra for $\pi^+$ in Pb+Pb collisions at LHC energies and compared our model predictions with the AMPT model results. With this comparison we found a reasonable agreement between the results of our thermal model with flow and that observed in the AMPT model. This validates the approach of our thermal model with the longitudinal flow in studying the rapidity spectra at LHC energies. We presented the variations of rapidity distributions for $\pi^+$ over a wide energy range from 8.7 GeV to 5.02 TeV with respect to the shifted rapidity $y-y_{beam}$. We found that rapidity distributions show the property of extended longitudinal scaling up to below top RHIC energy, while it is observed to be violated at RHIC 200 GeV and LHC energies in our thermal model with flow. Pion multiplicity data, $dN/d\eta(y)$ in the forward rapidities or with an extended (pseudo)rapidity interval at LHC energies would be extremely useful to verify the hypothesis of limiting fragmentation at higher collision energies and thus would help in fine-tuning the theoretical models to describe the particle production in heavy-ion collisions.

\section*{ACKNOWLEDGEMENTS}
The authors acknowledge stimulating discussions with Dr. Prakhar Garg at the beginning of the work.


\begin{thebibliography}{99}

\bibitem{Singh:1993}
C. P. Singh, Phys. Rep.  {\bf236}, 147 (1993).


\bibitem{Singh:1992}
C. P. Singh, Int. J. Mod. Phys. A {\bf7}, 7185 (1992).

 
\bibitem{Satz:2000}
 H. Satz, Rep. Prog. Phys. {\bf 63}, 1511 (2000).


\bibitem{Muller:1995}
 B. Muller, Rep. Prog. Phys. {\bf 58}, 611 (1995).

 
\bibitem{Shuryak:1980}
 E. V. Shuryak, Phys. Rep. {\bf 61}, 71 (1980).


\bibitem{Letessier:2004}
J. Letessier and J. Rafelski, Hadrons and Quark-Gluon Plasma, Cambridge University Press, U.K. (2004).


%%%%%%%%%%%%%%%%%%%%%%%%%%%%%%%%%%%%%%

\bibitem{Landau:1953}
 L. D. Landau, Izv. Akad. Nauk. Sec. Fiz. {\bf 17}, 51 (1953).
 
 
 \bibitem{Bjorken:1983}
J. D. Bjorken, Phys. Rev. D {\bf 27}, 140 (1983).


\bibitem{Schnedermann:1993}
 E. Schnedermann, J. Sollfrank, and U. Heinz, Phys. Rev. C {\bf 48}, 2462 (1993).


\bibitem{Braun:1996}
 P. Braun-Munzinger $\it et\; al.$, Phys.Lett. B {\bf 365}, 1 (1996).
 
 
 \bibitem{Braun:1995}
 P. Braun-Munzinger $\it et\; al.$, Phys. Lett. B {\bf 344}, 43 (1995).


\bibitem{Schnedermann:1992}
 E. Schnedermann, and U. Heinz, Phys. Rev. Lett. {\bf 69}, 2908 (1992).


\bibitem{Feng:2011}
 S. Q. Feng, and Y. Zhong, Phys. Rev. C {\bf 83}, 034908 (2011).


\bibitem{Feng:2009}
 S. Q. Feng and X. B. Yuan, Sci. China Ser. G {\bf 52}, 198 (2009).


\bibitem{Hirano:2002}
 T. Hirano, K. Morita, S. Muroya, and C. Nonaka, Phys. Rev. C {\bf 65}, 061902 (2002).
 
 
 \bibitem{Morita:2002}
 K. Morita, S. Muroya, C. Nonaka, and T. Hirano, Phys. Rev C {\bf 66}, 054904 (2002).


\bibitem{Manninen:2011}
 J. Manninen, E. L. Bratkovskaya, W. Cassing, and O. Linnyk, Eur. Phys. J. C {\bf 71}, 1615 (2011).


\bibitem{Bass:1998}
 S. A. Bass $et\; al.$, Prog. Part. Nucl. Phys. {\bf 41}, 225 (1998).



\bibitem{Mayer:1997}
 U. Mayer and U. Heinz, Phys. Rev. C {\bf 56}, 439 (1997).
 
 
 \bibitem{Heinz:1999}
 U. Heinz, Nucl. Phys. A {\bf 661}, 140c (1999).



\bibitem{Becattini:2007}
 F. Becattini and J. Cleymans, J. Phys. G {\bf 34}, S959 (2007).
 
 
 \bibitem{Becattini1:2007}
 F. Becattini, J. Cleymans and J. Strumpfer, Proceeding of Science, (CPOD07), {\bf 012} (2007) [arXiv:0709.2599[hep-ph]].


\bibitem{Biedron:2007}
 B. Biedron, and W. Broniowski, Phys. Rev. C {\bf 75}, 054905 (2007).
 
 
 \bibitem{Broniowski:2008}
 W. Broniowski and B. Biedron, J. Phys. G {\bf 35}, 0440189 (2008).

 
\bibitem{Cleymans:2008} 
  J.~Cleymans, J.~Strumpfer and L.~Turko,
  %``Extended longitudinal scaling and the thermal model,''
  Phys.\ Rev.\ C {\bf 78}, 017901 (2008).

\bibitem{Broniowski:2001}
 W. Broniowski, and W. Florkowski, Phys. Rev. Lett. {\bf 87}, 272302 (2001).


\bibitem{Broniowski:2002}
 W. Broniowski, and W. Florkowski, Phys. Rev. C {\bf 65}, 064905 (2002).


\bibitem{Benecke:1969}
J. Benecke, T. T. Chou, C. N. Yang, and E. Yen, Phys. Rev. {\bf 188}, 2159 (1969). 


\bibitem{Chou:1970}
T. T. Chou and C. N. Yang, Phys. Rev. Lett. {\bf 25}, 1072 (1970).


\bibitem{Feynman:1969}
R. P. Feynman, Phys. Rev. Lett. {\bf 23} , 1415 (1969).

\bibitem{Hagedorn:1970}
R. Hagedorn, Nucl. Phys. B {\bf 24}, 93 (1970).

%%%%%%%%%%%%%%%%%%%%%%%%%%%%%%%%%%%%%%%%%%%%%%%%%%%%%%%%%


%\bibitem{aleph} H. Stenzel, (ALEPH Collaboration), contributed paper to {\bf ICHEP2000} (2000). 


\bibitem{Back:2006} 
B. B. Back {\it et al.} (PHOBOS Collaboration), Phys. Rev. C {\bf 74}, 021902(R) (2006).


\bibitem{UA5} G. Alner {\it et al.}, (UA5 Collaboration), Z. Phys. C {\bf 33}, 1 (1986).

\bibitem{Alver:2009} 
B. Alver {\it et al.} (PHOBOS Collaboration), Phys. Rev. Lett. {\bf 102}, 142301 (2009).


\bibitem{Adams:2005aa} 
  J.~Adams {\it {\it et al.}} (STAR Collaboration),
  %``Multiplicity and pseudorapidity distributions of photons in Au + Au collisions at s(NN)**(1/2) = 62.4-GeV,''
  Phys.\ Rev.\ Lett.\  {\bf 95}, 062301 (2005).
  
  
  
\bibitem{Adams:2005cy} 
  J.~Adams {\it et al.} (STAR Collaboration),
  %``Multiplicity and pseudorapidity distributions of charged particles and photons at forward pseudorapidity in Au + Au collisions at s(NN)**(1/2) = 62.4-GeV,''
  Phys.\ Rev.\ C {\bf 73}, 034906 (2006).  


\bibitem{Abelev:2010} 
B. I. Abelev {\it et al.} (STAR Collaboration), Nucl. Phys. A {\bf 832}, 134 (2010).



\bibitem{BackR:2005} 
B. B. Back {\it et al.} (PHOBOS Collaboration), Phys. Rev. C {\bf 72}, 031901(R) (2005).
 

\bibitem{phobos-prl-91} 
B. B. Back {\it et al.}(PHOBOS Collaboration), Phys. Rev. Lett. {\bf 91}, 052303 (2003).



\bibitem{Adriani:2015iwv} 
 O.~Adriani {\it et al.} [LHCf Collaboration],
  %``Measurements of longitudinal and transverse momentum distributions for neutral pions in the forward-rapidity region with the LHCf detector,''
  Phys.\ Rev.\ D {\bf 94}, no. 3, 032007 (2016).


\bibitem{ALICE:2014rma} 
  B.~B.~Abelev {\it et al.} (ALICE Collaboration),
  %``Inclusive photon production at forward rapidities in proton-proton collisions at $\sqrt{s}$ = 0.9, 2.76 and 7 TeV,''
  Eur.\ Phys.\ J.\ C {\bf 75}, 146 (2015).
  
\bibitem{bialas} 
A. Bialas and M. Jezabek, Phys. Letts. B {\bf 590}, 233 (2004). 

\bibitem{Ruan:2010} 
  J.~Ruan and W.~Zhu,
  %``Particle multiplicities at LHC and deviations from limiting fragmentation,''
  Phys.\ Rev.\ C {\bf 81}, 055210 (2010).

\bibitem{Nasim:2011}
  M.~Nasim, C.~Jena, L.~Kumar, P.~K.~Netrakanti and B.~Mohanty,
  %``Longitudinal scaling of observables in heavy-ion collision models,''
  Phys.\ Rev.\ C {\bf 83}, 054902 (2011).


\bibitem{Gelis:2006} 
  F.~Gelis, A.~M.~Stasto and R.~Venugopalan,
  %``Limiting fragmentation in hadron-hadron collisions at high energies,''
  Eur.\ Phys.\ J.\ C {\bf 48}, 489 (2006).
  

\bibitem{Bleibel:2016} 
  J.~Bleibel, L.~V.~Bravina, A.~B.~Kaidalov and E.~E.~Zabrodin,
  %``How many of the scaling trends in $pp$ collisions will be violated at sqrt{s_NN} = 14 TeV ? - Predictions from Monte Carlo quark-gluon string model,''
  Phys.\ Rev.\ D {\bf 93}, 114012 (2016).


\bibitem{Stasto:2011zza} 
  A.~Stasto,
  %``Limiting fragmentation in hadronic collisions,''
  Nucl.\ Phys.\ A {\bf 854}, 64 (2011).


\bibitem{atlas-cross} 
G. Aad {\it et al.} (ATLAS Collaboration), Nature Commun. {\bf 2}, 463 (2011).


\bibitem{Back:2005} 
  B.~B.~Back {\it et al.},
  %``The PHOBOS perspective on discoveries at RHIC,''
  Nucl.\ Phys.\ A {\bf 757}, 28 (2005).
  

\bibitem{Edward:2016} 
  E.~K.~G.~Sarkisyan, A.~N.~Mishra, R.~Sahoo and A.~S.~Sakharov,
  %``Multihadron production dynamics exploring the energy balance in hadronic and nuclear collisions,''
  Phys.\ Rev.\ D {\bf 93}, 054046 (2016);
  Addendum: [Phys.\ Rev.\ D {\bf 93}, 079904 (2016)].


\bibitem{Bopp:2005cr} 
  F.~W.~Bopp, J.~Ranft, R.~Engel and S.~Roesler,
  %``Antiparticle to Particle Production Ratios in Hadron-Hadron and d-Au Collisions in the DPMJET-III Monte Carlo,''
  Phys.\ Rev.\ C {\bf 77}, 014904 (2008).
  
  
  \bibitem{Engel:1996yb} 
  R.~Engel, J.~Ranft and S.~Roesler,
  %``Photoproduction off nuclei and point - like photon interactions 1. Cross-sections and nuclear shadowing,''
  Phys.\ Rev.\ D {\bf 55}, 6957 (1997).
  
  
  \bibitem{Ostapchenko:2004ss} 
  S.~Ostapchenko,
  %``QGSJET-II: Towards reliable description of very high energy hadronic interactions,''
  Nucl.\ Phys.\ Proc.\ Suppl.\  {\bf 151}, 143 (2006).


\bibitem{Tiwari:2013} 
  S.~K.~Tiwari, P.~K.~Srivastava and C.~P.~Singh,
  %``The effect of flow on Hadronic Spectra in an Excluded-Volume Model,''
  J.\ Phys.\ G {\bf 40}, 045102 (2013).

\bibitem{Tiwari:2012} 
  S.~K.~Tiwari, P.~K.~Srivastava and C.~P.~Singh,
  %``Description of Hot and Dense Hadron Gas Properties in a New Excluded-Volume model,''
  Phys.\ Rev.\ C {\bf 85}, 014908 (2012).
  
  
  
 \bibitem{Tiwari:2013wga} 
  S.~K.~Tiwari and C.~P.~Singh,
  %``Particle production in Ultra-relativistic Heavy-Ion Collisions : A Statistical-Thermal Model Review,''
  Adv.\ High Energy Phys.\  {\bf 2013}, 805413 (2013). 
  
  
\bibitem{Mishra:2008tc} 
  M.~Mishra and C.~P.~Singh,
  %``Particle multiplicities and particle ratios in excluded volume model,''
  Phys.\ Rev.\ C {\bf 78}, 024910 (2008).


\bibitem{Tiwari:2013pva} 
  S.~K.~Tiwari and C.~P.~Singh,
  %``Production of Strange, Non-strange particles and Hypernuclei in an Excluded-Volume Model,''
  J.\ Phys.\ Conf.\ Ser.\  {\bf 509}, 012097 (2014).


\bibitem{Netrakanti:2005}
 P. K. Netrakanti, and B. Mohanty, Phys. Rev. C {\bf 71}, 047901 (2005).


\bibitem{Abelev:2012wca} 
  B.~Abelev {\it et al.} [ALICE Collaboration],
  %``Pion, Kaon, and Proton Production in Central Pb--Pb Collisions at $\sqrt{s_{NN}} = 2.76$ TeV,''
  Phys.\ Rev.\ Lett.\  {\bf 109}, 252301 (2012)


\bibitem{Steinheimer:2013} 
  J.~Steinheimer, J.~Aichelin and M.~Bleicher,
  %``Nonthermal p/? Ratio at LHC as a Consequence of Hadronic Final State Interactions,''
  Phys.\ Rev.\ Lett.\  {\bf 110}, 042501 (2013).
  
  

\bibitem{Noronha:2014} 
  J.~Noronha-Hostler and C.~Greiner,
  %``Understanding the $p/\pi$ ratio at LHC due to QCD mass spectrum,''
  Nucl.\ Phys.\ A {\bf 931}, 1108 (2014).

%-----------------
\bibitem{Song-Bass} 
H. Song, S. Bass and U. Heinz, Phys. Rev. C {\bf 83}, 024912 (2011).



  
  
\bibitem{Abelev:2013vea} 
  B.~Abelev {\it et al.} (ALICE Collaboration),
  %``Centrality dependence of $\pi$, K, p production in Pb-Pb collisions at $\sqrt{s_{NN}}$ = 2.76 TeV,''
  Phys.\ Rev.\ C {\bf 88}, 044910 (2013).



\bibitem{Ma:2016fve}  G.~L.~Ma and Z.~W.~Lin,
  %``Predictions for $\sqrt {s_{NN}}=5.02$ TeV Pb+Pb Collisions from a Multi-Phase Transport Model,''
  Phys.\ Rev.\ C {\bf 93}, 054911 (2016).
  

\bibitem{Nayak:2016} 
  S.~Basu, T.~K.~Nayak and K.~Datta,
  %``Beam energy dependence of pseudorapidity distributions of charged particles produced in relativistic heavy-ion collisions,''
  Phys.\ Rev.\ C {\bf 93}, 064902 (2016).


\bibitem{Adamova:2003}
 D. Adamova $ \it et\; al.$, (CERES Collaboration), Phys. Rev. Lett. {\bf 90}, 022301 (2003).


\bibitem{Braun-Munzinger:2014lba} 
  P.~Braun-Munzinger, A.~Kalweit, K.~Redlich and J.~Stachel,
  %``Confronting fluctuations of conserved charges in central nuclear collisions at the LHC with predictions from Lattice QCD,''
  Phys.\ Lett.\ B {\bf 747}, 292 (2015).

 
 \end{thebibliography}
\end{document}